\documentclass[
showpacs,preprintnumbers,
nofootinbib,
 amsmath,amssymb,
 aps,
prd,
twocolumn
]{revtex4-1}

\usepackage[utf8]{inputenc}
\usepackage{epsf}
\usepackage{graphicx} 
\usepackage{amssymb}
\usepackage{dcolumn}
\usepackage{amsmath}
\usepackage{hyperref}
\usepackage{multirow}
\usepackage{booktabs}
\usepackage{mathtools}
\usepackage{siunitx}
\usepackage{soul}
\usepackage{color}

\def\thetaB{\mbox{\boldmath$\theta$}}

\def\rB{\mbox{\boldmath$r$}}
\def\ellB{\mbox{\boldmath$\ell$}}

\def\lsim{~\rlap{$<$}{\lower 1.0ex\hbox{$\sim$}}}
\def\gsim{~\rlap{$>$}{\lower 1.0ex\hbox{$\sim$}}}

\def\cn{\mathcal{C}}
\def\kp{\kappa_{\rm proj}}
\def\kl{\kappa_{\rm lens}}
\def\rv{r_{\rm vir}}
\def\Mv{M_{\rm vir}}

\makeatletter
\newcommand{\pushright}[1]{\ifmeasuring@#1\else\omit\hfill$\displaystyle#1$\fi\ignorespaces}
\newcommand{\pushleft}[1]{\ifmeasuring@#1\else\omit$\displaystyle#1$\hfill\fi\ignorespaces}
\makeatother

\begin{document}

\title{The Origin of Weak Lensing Convergence Peaks}
\author{Jia Liu$^{1,2}$}
 \email{jia@astro.princeton.edu}
 \thanks{NSF Astronomy and Astrophysics Postdoctoral Fellow}
 \author{Zolt\'an Haiman$^2$} 
\affiliation{ $^1$ Department of Astrophysical Sciences, Princeton University, Princeton, NJ 08544, USA}  
\affiliation{ $^2$ Department of Astronomy, Columbia University, New York, NY 10027, USA} 

\date{\today}
\begin{abstract}
Weak lensing convergence peaks are a promising tool to probe nonlinear structure evolution at late times, providing additional cosmological information beyond second-order statistics. Previous theoretical and observational studies have shown that the cosmological constraints on $\Omega_m$ and $\sigma_8$ are improved by a factor of up to $\approx$ 2 when peak counts and second-order statistics are combined, compared to using the latter alone. We study the origin of lensing peaks using observational data from the 154 deg$^2$ Canada-France-Hawaii Telescope Lensing Survey. We found that while high peaks (with height $\kappa>3.5 \sigma_\kappa$, where $\sigma_\kappa$ is the r.m.s. of the convergence $\kappa$) are typically due to one single massive halo of $\approx10^{15}M_\odot$, low peaks ($\kappa\lsim\sigma_\kappa$) are associated with constellations of 2--8 smaller halos ($\lsim10^{13}M_\odot$). In addition, halos responsible for forming low peaks are found to be significantly offset from the line-of-sight towards the peak center (impact parameter $\gsim$ their virial radii), compared with $\approx 0.25$ virial radii for halos linked with high peaks, hinting that low peaks are more immune to baryonic processes whose impact is confined to the inner regions of the dark matter halos. Our findings are in good agreement with results from the simulation work by Yang el al. 2011~\cite{Yang2011}.
\end{abstract}
\maketitle

\section{Introduction}

Weak lensing describes the effect of bending of background light rays by foreground matter (see~\cite{Kilbinger2015} for a recent review). It is sensitive to the large scale structure of the universe, and hence is a promising method to answer some unsolved fundamental questions in physics, such as the nature of dark energy and the total mass of neutrinos. Within the next decade, lensing datasets of unprecedented precision will come online from large surveys, such as the Euclid mission\footnote{\url{http://sci.esa.int/euclid}}, the Wide-Field Infrared Survey Telescope\footnote{\url{http://wfirst.gsfc.nasa.gov}}, and the Large Synoptic Survey Telescope\footnote{\url{http://www.lsst.org}}. These surveys will be sensitive to structure evolution in the strongly nonlinear regime, warranting a study of higher-order (non-Gaussian) statistics that contain information beyond the traditional second-order statistics, such as the two-point correlation function and its Fourier transform, the power spectrum. 

The abundance of peaks in lensing convergence maps has been proposed to be a powerful tool to probe non-Gaussianity in the density field~\cite{Jain2000,JV00}. Peak counts describe the distribution of local maxima (i.e. pixels with values higher than their eight neighbors) in a convergence map as a function of their heights. It is a particularly simple statistic, forecasted by several theoretical studies~\cite{Dietrich2010,Kratochvil2010,Marian2011,Marian2012,Yang2011,Yang2013} to yield
a factor of $\approx$ two improvement on cosmological parameters when combined with second-order statistics. Recently, cosmological constraints from peak counts have been obtained using data from the 154 deg$^2$ Canada-France-Hawaii Telescope Lensing Survey (CFHTLenS) \cite{Liu2015}, the 130 deg$^2$ CFHT-Stripe 82 survey~\cite{Liu2015b}, and the 139 deg$^2$ Dark Energy Survey (DES) Science Verification run~\cite{Kacprzak2016}. All results are consistent with constraints from second-order statistics. The error contour from peaks alone is comparable in size with that from second-order statistics, and a factor of two improvement is seen when they are combined. While Refs.~\cite{Liu2015, Kacprzak2016} use full N-body simulations for modeling peaks, Ref.~\cite{Liu2015b} adopts a halo-based model~\cite{Fan2010}. 

The goal of this paper is to understand the physical origin of convergence peaks, using observational data from  CFHTLenS. Semi-analytical studies of peaks by various groups~\cite{Fan2010,Lin&Kilbinger2015a,Lin&Kilbinger2015b,Lin2016} all rely heavily on the assumption that peaks are projections of foreground dark matter halos. Motivated by these theoretical works, we here search for the peak--halo connection in observations. Previously, Ref.~\cite{Yang2011} (hereafter Y11) has conducted a such study with simulations, where they traced individual peaks back into the N-body box and searched for  halos along or close to the line-of-sight. They found that high peaks ($\kappa_{\rm peak}>3.5\sigma_\kappa$, where $\sigma_\kappa$ is the r.m.s. of the convergence map) are normally linked to one single massive halo, while low peaks ($\kappa_{\rm peak}\lsim 1.5\sigma_\kappa $) are associated with several (typically 4--8) smaller halos within $\approx 1$ arcmin of the line-of-sight. These halos are in the mass range of $10^{12}$--$10^{13}M_\odot$, which we expect to host galaxies well above the brightness detection threshold of CFHTLenS. 
In their follow-up work~\cite{Yang2013} they showed that, while high peaks are sensitive to baryonic processes, low peaks are relatively immune to baryons, as long as these processes are confined to the core of the halo. The latter peaks are typically offset from the center of contributing halos by $\approx$ the virial radius, a distance at which baryons are expected to have much less impact on the matter distribution than near the cores of halos (\cite{Schaller2015}, but see~\cite{Semboloni2011}).

This paper is organized as follows. In Sec.~\ref{data}, we describe the creation of our lensing maps $\kl$ and projected mass maps $\kp$. We then demonstrate the connection between lensing peaks and halos, identify halos that contribute to peaks, and study their properties in Sec.~\ref{results}. Finally, we conclude in Sec.~\ref{conclusion}.

\section{Data}\label{data}

From the CFHTLenS catalogue, we produce two types of convergence maps: (1) $\kl$ maps from shear measurements, and (2) $\kp$ maps from projected foreground halos, with mass $M_{\rm vir}$ estimated from the stellar mass. We first show the close connection between these two sets of maps, and then study the properties of contributing halos using information stored during the $\kp$ map creation.

The CFHTLenS survey consists of four sky patches, with a total area of 154 deg$^2$ and a limiting magnitude of $i_{\rm AB}\lesssim24.5$. Each galaxy in the catalogue has  photometric redshift, ellipticity, and stellar mass measurements. An overview of the data reduction procedure is detailed in Appendix C of Ref.~\cite{Erben2013}, with more in-depth analyses presented in Refs.~\cite{Hildebrandt2012,Heymans2012,Miller2013,Kilbinger2013}. Throughout this work, we use  best-fit cosmological parameters inferred from the cosmic microwave background anisotropies by Planck (``TT, TE, EE+lowP'' in Table 4 of \cite{Planck2015XIII}).

\subsection{Convergence Map from Shear Measurements ($\kl$)}

\begin{figure*}
\begin{center}
\includegraphics[width=0.9\textwidth]{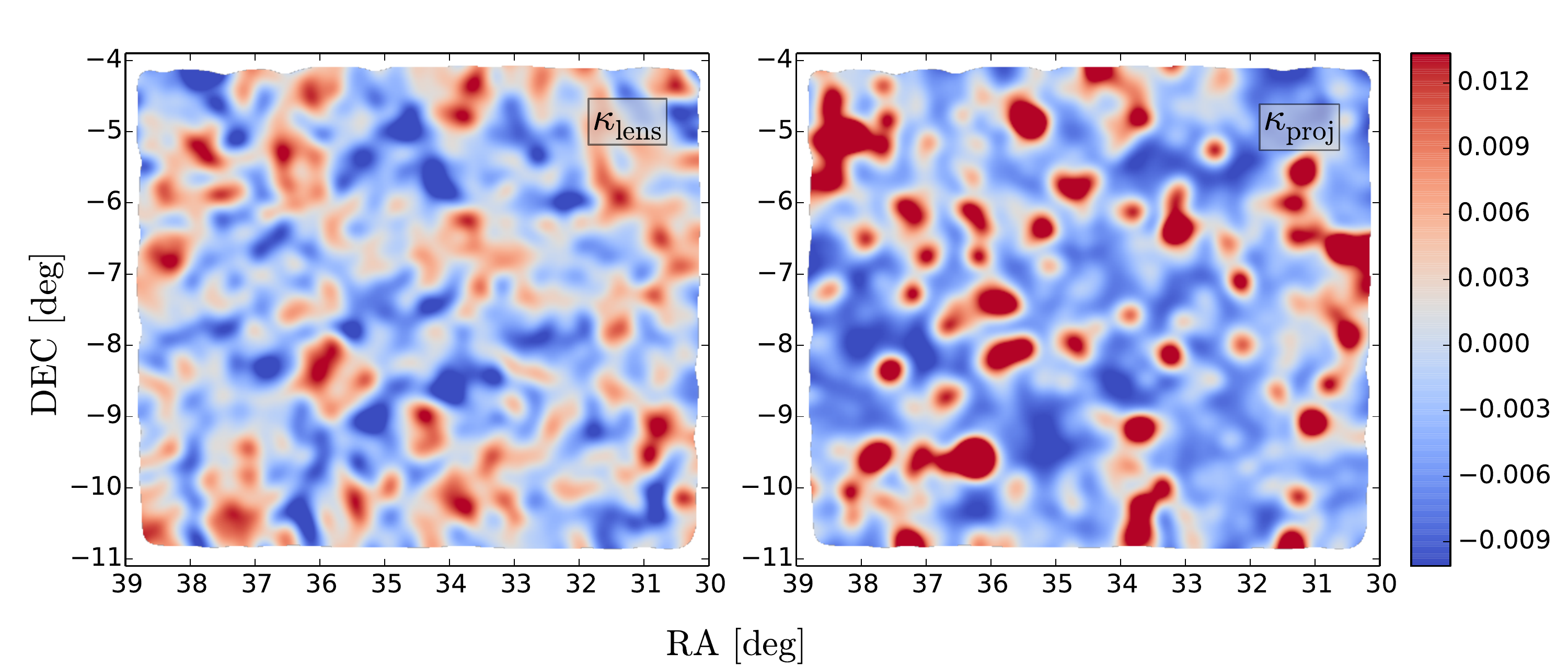}
\end{center}
\caption{\label{fig:conv_W1} Reconstructed $\kl$ map from shear measurements (left) and $\kp$ map from projected foreground halos (right), for the CFHTLenS W1 field. Both maps are smoothed with an 8.9 arcmin Gaussian window.} 
\end{figure*}

To select background sources, we applied the following cuts to the CFHTLenS catalogue: redshift $0.4<z<1.3$,  \verb star_flag $\;= 0$ (to exclude stars),  \verb lensfit $\;$ weight~{\texttt w}~$> 0$ (to exclude galaxies with failed shape measurements), 
and {\texttt mask} $\leq 1$ (to exclude galaxies falling within large masks around stars, asteroid trails, or bad pixels). We obtain 4.5 million galaxies, with a mean redshift $\langle z\rangle$ = 0.8.

We construct our lensing convergence maps using the method in Ref.~\cite{Kaiser1993},
\begin{eqnarray}
\label{eq: KSI}
\hat{\kappa}_{\rm lens}(\ellB) = \left(\frac{\ell_1^2-\ell_2^2}{\ell_1^2+\ell_2^2}\right)\hat{\gamma}_1(\ellB)
+ 2\left(\frac{\ell_1\ell_2}{\ell_1^2+\ell_2^2}\right)\hat{\gamma}_2(\ellB),
\end{eqnarray}
where $\hat{\kappa}_{\rm lens}, \hat{\gamma_1}$, and $\hat{\gamma_2}$ are the 
lensing convergence and components of the shear in Fourier space, and $\ellB$ is the wavevector 
with components $(\ell_1, \ell_2)$.  We use the observed ellipticity $e_1$ and $e_2$ as the estimators of $\gamma_1,\gamma_2$. Finally, applying the inverse Fourier transform on $\hat{\kappa}_{\rm lens}$, we obtain the convergence map in real space. As an example, we show the reconstructed CFHTLenS W1 field in the left panel of Fig.~\ref{fig:conv_W1}, smoothed with an 8.9 arcmin Gaussian window. To minimize the bias due to masking~\cite{LiuX2014b}, we discard peaks that fall within the masks, although we note that any residual biases should be the same in our simulated maps and the maps derived from the data, since the same mask and same procedure is applied to both.

\subsection{Projected Mass Map from Foreground Halos ($\kp$)}

Under the thin-lens approximation\footnote{In this limit, the size of the lens is much smaller than the distances between the source, lens, and observer.}, lensing convergence is defined to be the ratio of the line-of-sight projection of the mass density $\Sigma$ to the critical surface mass density $\Sigma_{\rm cr}$,
\begin{align}
\kp(\thetaB)=\frac{\Sigma(\thetaB)}{\Sigma_{\rm cr}}
\end{align}
with
\begin{align}
\Sigma(\thetaB)&=\int d\chi \; \rho(\chi, \rB=D\thetaB) \\
\Sigma_{\rm cr}&=\frac{c^2}{4\pi G}\frac{D_s}{D_{ls}D_l}
\end{align}
where $\chi$ is the comoving distance, $\rho$ is the halo density profile (we explicitly assume that all convergence contributions are from halos alone, and for simplicity, project the halo contributions without ray-tracing), $c$ is the speed of light, and $G$ is the gravitational constant. $D$ is the angular diameter distance, and $D_s$, $D_{ls}$, and $D_{l}$ denote that between the source and the observer, the lens and the source, and the lens and the observer, respectively. Assuming an Navarro-Frenk-White (NFW)~\cite{NFW1996} density profile for each halo, we replace $\rho$ by $\rho_{\rm nfw}$,
\begin{align}
\rho_{\rm nfw} (\chi, r=|\rB|) &= \frac{\rho_s}{x\left[1+x\right]^2},\, x:=\frac{r}{r_s}
\end{align}
where $\rho_s, r_s$ are the characteristic density and radius of the halo, we can obtain an analytical form of the projected mass density~\cite{Takada2003},
\begin{align}
\Sigma(\thetaB)=2\rho_s r_sG(x)
\end{align}
with the projection factor,
\begin{align}
 G(x)=
  \begin{dcases}
          \frac{\sqrt{\cn^2-x^2}}{(x^2-1)(1+\cn)}+\frac{1}{(1-x^2)^{3/2}}{\rm arcosh}\left[\frac{x^2+\cn}{x(1+\cn)}\right],\\
          \hskip0.34\textwidth(x<1)\\ 
          \frac{\sqrt{\cn^2-1}}{3(1+\cn)}\left(1+\frac{1}{\cn+1}\right),\hskip0.137\textwidth (x=1)\\
          \frac{\sqrt{\cn^2-x^2}}{(x^2-1)(1+\cn)}-\frac{1}{(x^2-1)^{3/2}}{\rm arccos}\left[\frac{x^2+\cn}{x(1+\cn)}\right],\\
          \hskip0.3\textwidth(1<x\le \cn)\\
          0, \hskip0.316\textwidth (x> \cn)\\
          \end{dcases} %
\end{align}
where $\cn \equiv \rv / r_s$ is the concentration parameter. The virial radius $\rv(z)$ is defined as the radius of a spherical volume within which the mean density is $\Delta_c$ times the critical density $\rho_c$ at redshift $z$,
\begin{align}
\rv(z)= \left[\frac{3}{4\pi} \frac{\Mv}{\Delta_c (z) \rho_c (z)} \right]^{\frac{1}{3}}.
\end{align}

We adopt a redshift dependent concentration parameter~\cite{Munoz-Cuartas2011},
\begin{align}
\log(\cn)&=a(z)\log(\Mv/[h^{-1}M_{\odot}])-b(z)\;,\\
a(z)&=\lambda z-\xi\;,\\
b(z)&=\frac{\mu}{(z+\zeta)}+\frac{\nu}{(z+\zeta)^2}\;,
\end{align}
where $\lambda=0.029$, $\xi=0.097$, $\mu=-110.001$, and $\nu=2469.720$. In addition,
 we use the fitting formula for $\Delta_c$ in Ref.~\cite{BryanNorman1998} (assuming a flat universe),
\begin{align}
\Delta_c &= 18\pi^2+82d-39d^2, \\
d&=\frac{\Omega_m(1+z)^3}{\Omega_m(1+z)^3+\Omega_\Lambda}-1.
\end{align}

At the location of each CFHTLenS galaxy, we assign a halo using the empirical stellar-to-halo mass relation from Ref.~\cite{Leauthaud2012},
\begin{align}
\log_{10} \left(\frac{\Mv}{M_{\rm vir,0}}\right)= \beta \log_{10} \left(\frac{M_*}{M_{*,0}}\right)
+\frac{\left(\frac{M_*}{M_{*,0}}\right)^\delta}{1+\left( \frac{M_*}{M_{*,0}} \right)^{-\gamma}} - \frac{1}{2}
\end{align}
where $M_{\rm vir,0}$ and  $M_{*,0}$ are the characteristic halo and stellar mass, $\beta$ and $\delta$ are the low and high-mass slopes, respectively, with a transitional parameter $\gamma$. For these five parameters, we use the redshift-dependent best-fits in Table~5 (\texttt{SIG MOD1} model) of Ref.~\cite{Leauthaud2012}. 

With the formalism described above, we can construct a set of projected mass maps ($\kp$) using the stellar mass information in the CFHTLenS catalogue. We use exactly the same sources as in the $\kl$ maps ($0.4<z<1.3$), but extend the lower bound of the redshift to $z=0.2$ for the lenses ($0.2<z<1.3$). Although the source galaxies will inevitably be lensed by foreground halos at $z<0.2$, we excluded these lenses as their photometric redshift measurements are not accurate. We do not expect this to significantly impact our results, as there is only a small number of very low redshift galaxies, and they reside at the tail of the lensing efficiency kernel. In addition, we also allow lenses to have $\texttt w=0$---these galaxies are typically too large to be enclosed in the postage stamps used to measure galaxy shapes, hence are excluded from the source sample. They are likely to be bright massive galaxies that contribute significantly to the lensing signal, and therefore must be included as lenses. 

For individual sources, we sum up the convergence contribution from all foreground halos that are within a 20 arcmin radius and at a lower redshift\footnote{We choose a 20 arcmin search radius so that it is large enough to account for all possible contributing halos. We later found that 10 arcmin is sufficient, and hence reduce the search radius accordingly in Sec.~\ref{subsec: peak-halo}.}. We then apply the same weight ${\texttt w} $ and smoothing kernel as done for the $\kl$ maps.  We here also assume the Born approximation, where the light-ray path is approximated as a straight line. This effectively induces a second-order noise to the $\kp$ maps. We expect this to be a subdominant effect at the current level of uncertainties in our halo mass estimation. However, the impact of the Born approximation on convergence peaks should be quantified in the future.

For a comparison, we show both the $\kl$ and the $\kp$ maps for the W1 field in Fig.~\ref{fig:conv_W1}. Coherent structures are clearly seen in both maps. The differences between the two maps are likely due to the shape noise in the $\kl$ map and the scatter in the stellar-to-halo mass relation related to the $\kp$ map. We note that $\kp$ does not include contributions from undetected halos, likely less massive halos, as well as non-halo structures such as filaments and walls.

\subsection{Correlation between $\kl$ \& $\kp$ maps}

\begin{figure*}
\begin{center}
\includegraphics[width=\textwidth]{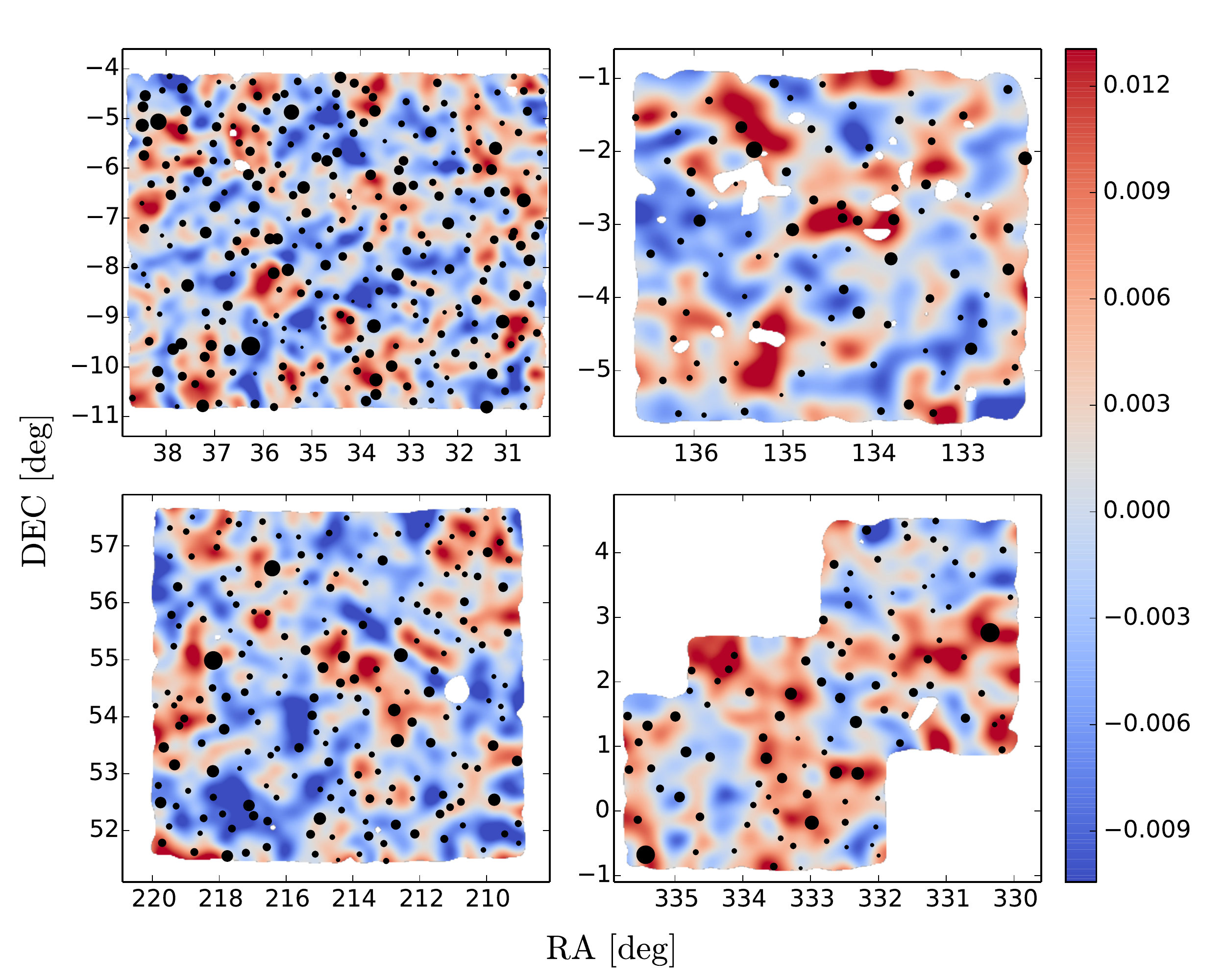}
\end{center}
\caption{\label{fig:matching_peaks} Peaks (black filled circles) found in $\kp$ maps are overlaid on $\kl$ maps, for CFHTLenS W1, W2, W3, W4 fields (upper-left, upper-right, lower-left, and lower-right, respectively). Circle sizes represent the height of peaks. $\kp$ maps are smoothed with a 5.3 arcmin Gaussian window, while $\kl$ maps with a 8.9 arcmin window. These two different smoothing scales are used for the purpose of visual display. } 
\end{figure*}

\begin{figure}
\begin{center}
\includegraphics[width=0.47\textwidth]{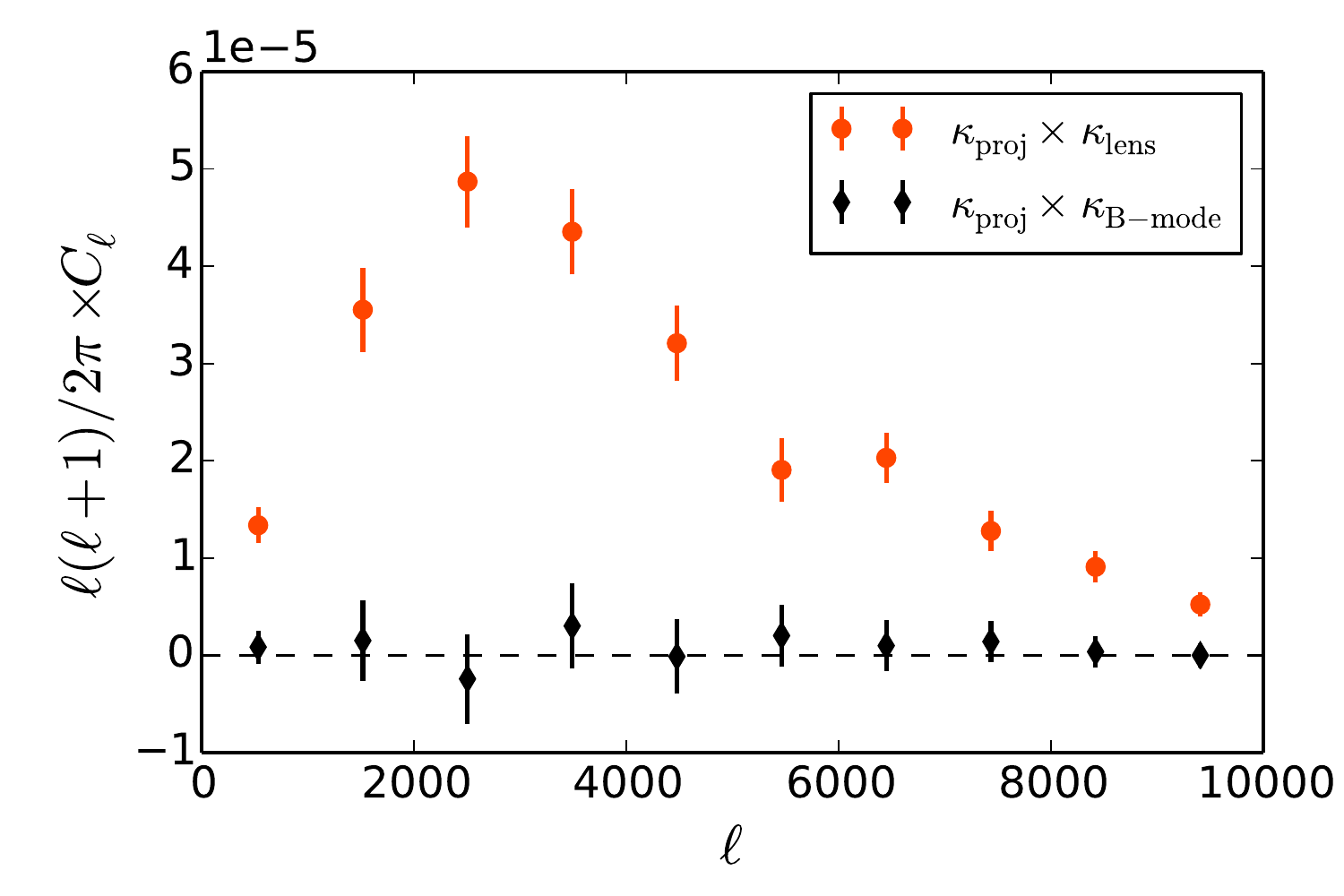}
\end{center}
\caption{\label{fig:CC} Cross-correlations of the $\kp$ and $\kl$ maps (red circle) and of the $\kp$ and $\kappa_{\rm B-mode}$ maps (black diamonds), with analytical Gaussian error estimates. The correlation between the $\kp$ and $\kl$ maps has a significance of 24.1.} 
\end{figure}

In Fig.~\ref{fig:matching_peaks}, we overlay peaks found in $\kp$ maps  
on top of $\kl$ maps (four panels for four CFHTLenS fields). 
$\kp$ peaks can be seen to roughly trace out structures in $\kl$ maps.  To validate our map making procedure, we cross-correlate the $\kl$ and $\kp$ maps. We first obtain the two-dimensional cross-power spectrum,
\begin{align}
\label{eq: ps2d}
C^{\kl \kp}(\ellB) = \hat \kappa_{\rm lens} (\ellB)^*\hat  \kappa_{\rm proj}(\ellB) \,,
\end{align}
where $*$ denotes complex conjugation. We then average over pixels in each multipole bin, $|\ellB| \in (\ell-\Delta\ell/2, \, \ell+\Delta\ell/2)$, for 10 linearly spaced bins between $40\leq\ell\leq10,000$, to obtain the one-dimensional power spectrum $C_\ell$. We use the Gaussian error (e.g., see Eq. 3 in Ref.~\cite{Planck2013XVIII}) as a rough estimate of the significance of the correlation,
\begin{align}
\left(\Delta C_\ell^{\kl \kp}\right)^2=\frac{C_\ell^{\kl \kl} C_\ell^{\kp \kp} + \left(C_\ell^{\kl \kp}\right)^2}{f_{\rm sky}(2\ell+1)\Delta\ell},
\end{align}
\begin{align}
{\rm SNR} = \sqrt{\sum_\ell \left(\frac{C_\ell^{\kl \kp}}{\Delta C_\ell^{\kl \kp}}\right)^2}
\end{align}
where $f_{\rm sky}=0.0034$ is the fraction of sky coverage. We weigh each of the four CFHTLenS fields by the inverse of their variance. Our results are shown in Fig.~\ref{fig:CC}, where a clear correlation is seen, with a signal-to-noise ratio (SNR) of 24.1. 

For a null test, we create a set of lensing B-mode maps ($\kappa_{\rm B-mode}$) by rotating galaxies in $\kl$ maps by 45$^\circ$. We expect $\kappa_{\rm B-mode}$ to be pure noise at the current level of detection, hence no correlation between $\kp$ and $\kappa_{\rm B-mode}$ maps. The null test results are also shown in Fig.~\ref{fig:CC}, where the correlation is consistent with 0.

\section{Results}\label{results}

\subsection{Cosmological signals in $\kl$ peaks}

To inspect the cosmological information stored in peaks, we compare peak counts from $\kl$ maps with that from a set of Gaussian random fields. To produce Gaussian random maps, we rotate each galaxy in the catalogue by a random angle, while keeping the amplitude of their ellipticity. We then follow the same procedure as in Eq.~\ref{eq: KSI} to create $\kappa_{\rm random}$ maps. In total, we create 500 realizations. These maps contain pure noise from galaxy shapes, hence any deviation from it is a sign of potential cosmological sensitivity. 

\begin{figure}
\begin{center}
\includegraphics[width=0.49\textwidth]{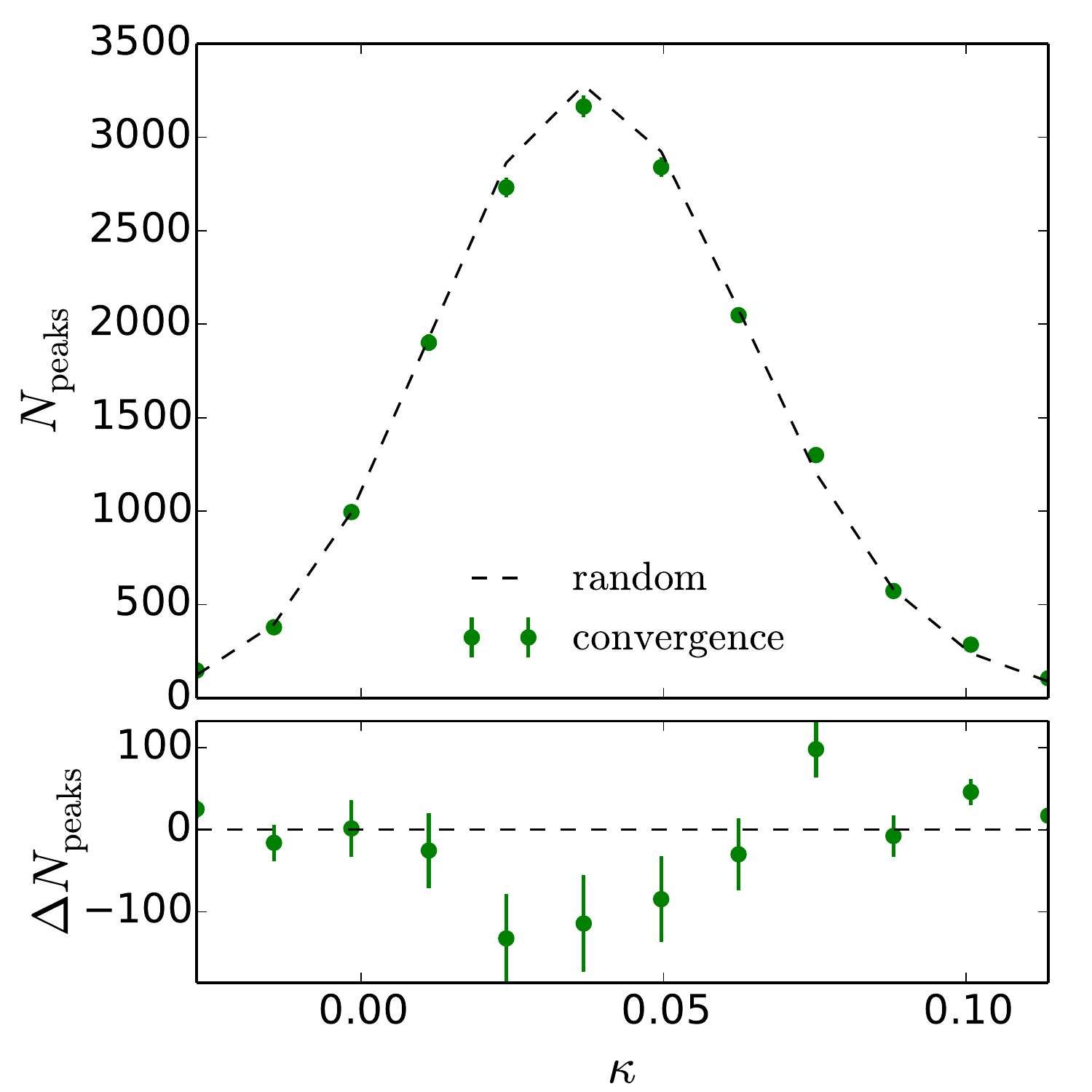}
\end{center}
\caption{\label{fig:kl_peaks} The top panel shows peak counts from $\kl$ maps (filled green circles), compared with averaged peak counts from 500 random noise maps (dashed curves), created by rotating galaxies by a random angle. The bottom panel shows their difference. The error bars are the standard deviation on 500 random realizations. Maps are smoothed with a 1 arcmin Gaussian window and have r.m.s. of $\sigma_{\kappa}=0.036$. $\kl$ peak counts deviate from the random case significantly (SNR=6.2).} 
\end{figure}

We show peak counts from $\kl$ and $\kappa_{\rm random}$  maps in Fig.~\ref{fig:kl_peaks}, both smoothed with a 1.0 arcmin Gaussian window, which is close to the optimal filter size  for peak counts~\cite{Liu2015}.  We detect a significant deviation (SNR=6.2) of $\kl$ peak counts from $\kappa_{\rm random}$ peak counts. At the medium--high significance region ($\kappa > 2\sigma_\kappa$, where $\sigma_\kappa=0.036$), we see an unsurprising excess of $\kl$ peaks, due to the highly nonlinear structure formation at late times. In contrast, we see a systematic deficit of  peak counts in the low significance region ($\kappa\approx 1 \sigma_{\kappa}$). This is consistent with predictions in Refs.~\cite{Kratochvil2010,Yang2011} and also peak counts measured in the DES survey (see Fig.3 in Ref.~\cite{Kacprzak2016}). Y11 has shown that there is rich cosmological information stored in these low peaks, despite the large contribution from noise to their heights, and that the constraining power of low peaks is comparable or better than that of  high peaks. Therefore, it is important to understand the origin of these low peaks, and eventually include them in peak count models. So far, only theoretical work using N-body simulations can incorporate low peaks, while simpler models~\cite{Fan2010,Lin&Kilbinger2015a,Lin&Kilbinger2015b,Lin2016} only focus on high peaks.

\subsection{Peak--halo connection}\label{subsec: peak-halo}

\begin{figure}
\begin{center}
\includegraphics[width=0.47\textwidth]{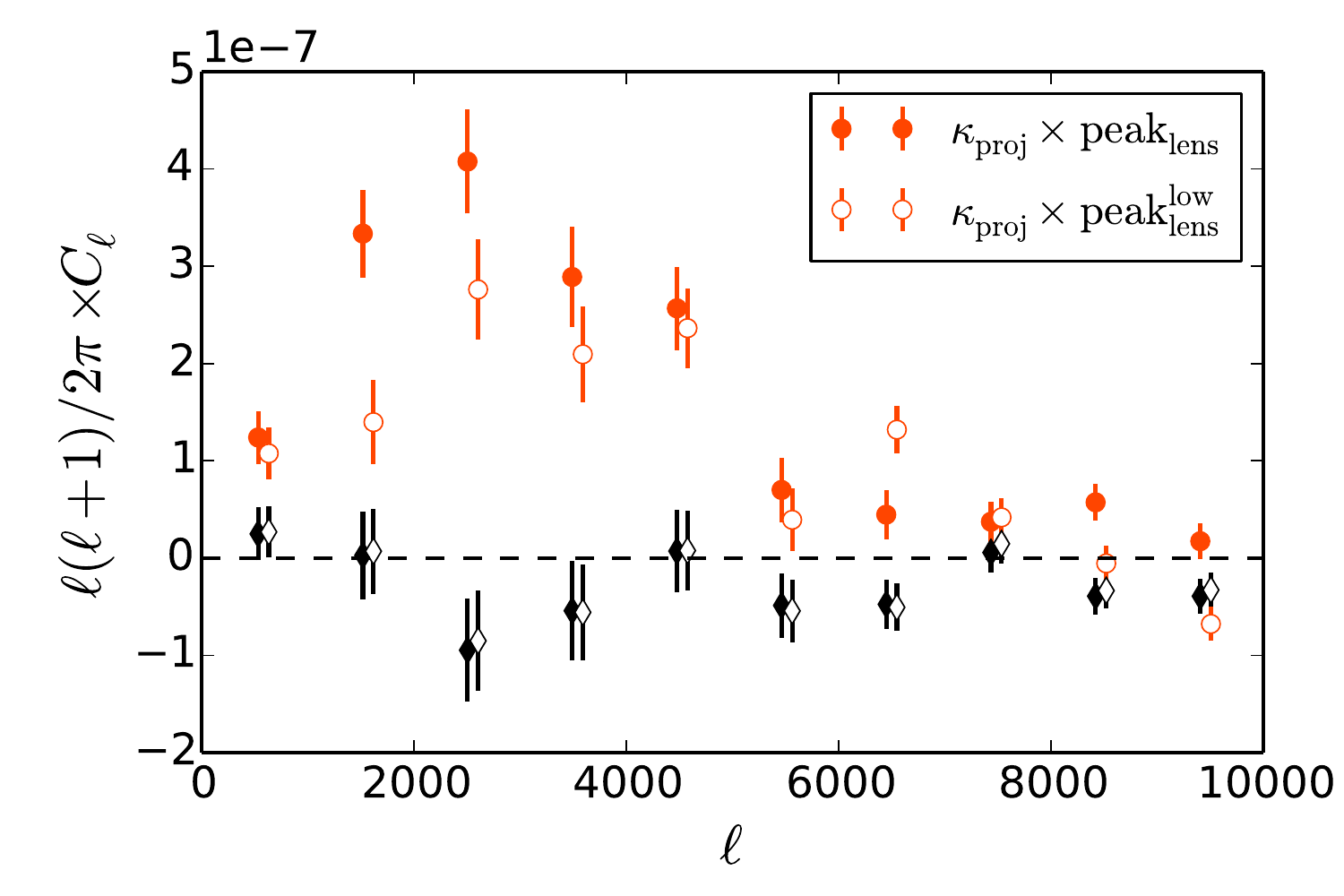}
\includegraphics[width=0.47\textwidth]{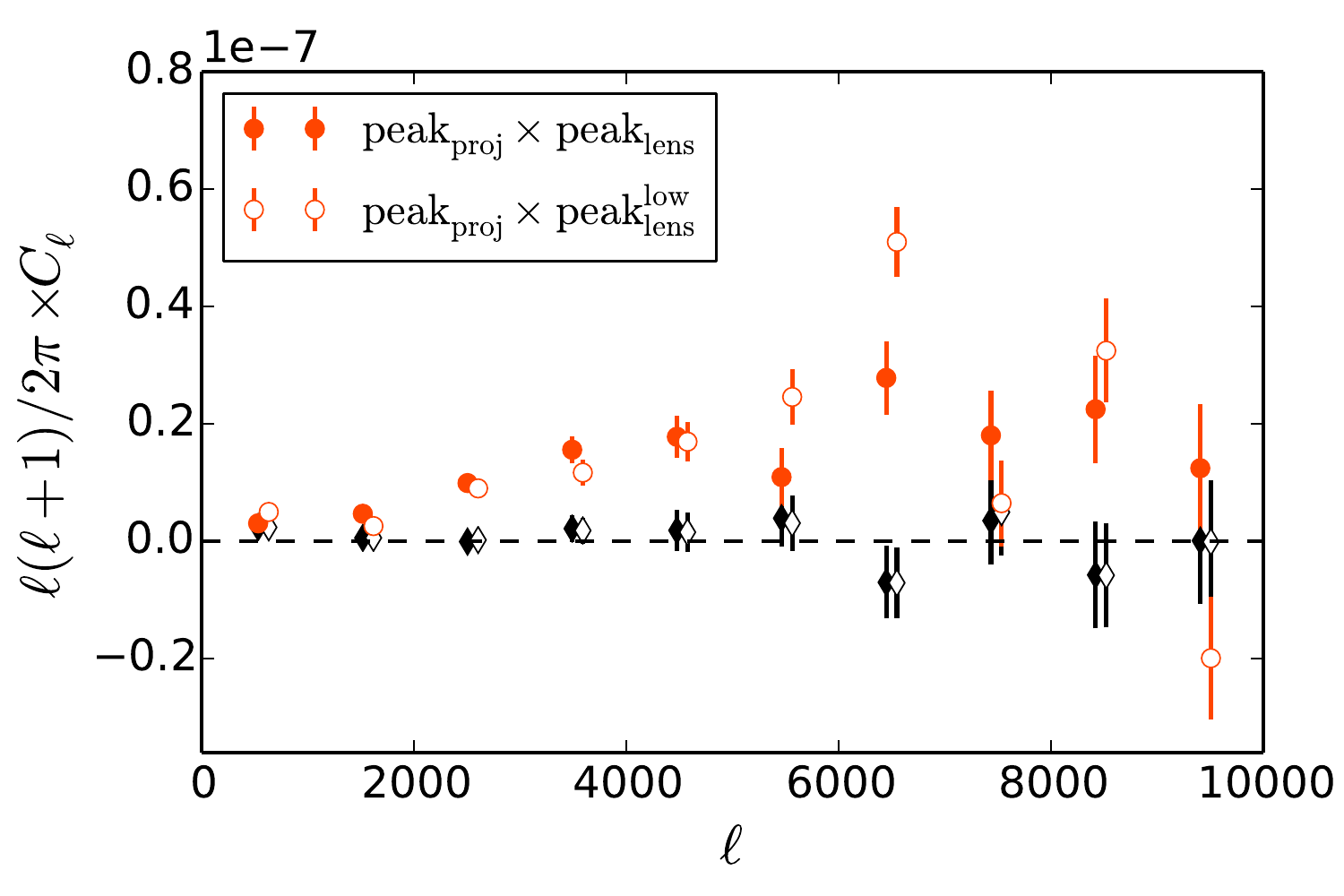}
\end{center}
\caption{\label{fig:CC_peaks} {\em Top panel}: cross-correlations of $\kp$ maps and $\kl$ peaks.  Filled red circles show the relation for all $\kl$ peaks (SNR=16.1). Open red circles include only low $\kl$ peaks ($<3.5\sigma_\kappa$), where the signal, though slightly reduced, remains significant (SNR=14.0). {\em Bottom panel}: cross-correlation of $\kp$ peaks and $\kl$ peaks, with SNR=23.7 (all peaks, red filled circle) and SNR=21.9 (low peaks only, red open circle). We also show results of null tests (filled and open black diamonds for all and low peaks, respectively), where we replace $\kl$ peaks by $\kappa_{\rm B-mode}$ peaks.  All maps are smoothed with an 1.0 arcmin Gaussian window. } 
\end{figure}

We demonstrate explicitly the connection between lensing peaks and halos in Fig.~\ref{fig:CC_peaks}, where we cross-correlate (1) $\kl$ peaks with the $\kp$ maps, and (2) $\kl$ peaks with $\kp$ peaks.  Peak maps have the same value as $\kappa$ maps at peak locations, but all other pixels set to zero. Since $\kp$ maps are created from projected halos, any significant correlation amounts to a proof of the peak--halo connection. Both cross-correlations confirm this assumption, with SNR=16.1 and 23.7, respectively. Furthermore, when we limit our sample to low $\kl$ peaks only ($<3.5\sigma_\kappa$), the signal remains significant, with SNR=14.0 and 21.9, respectively. This again matches findings by Y11---low peaks are just as important as high peaks, if not more, in constraining cosmological parameters.

\begin{figure}
\begin{center}
\includegraphics[width=0.49\textwidth]{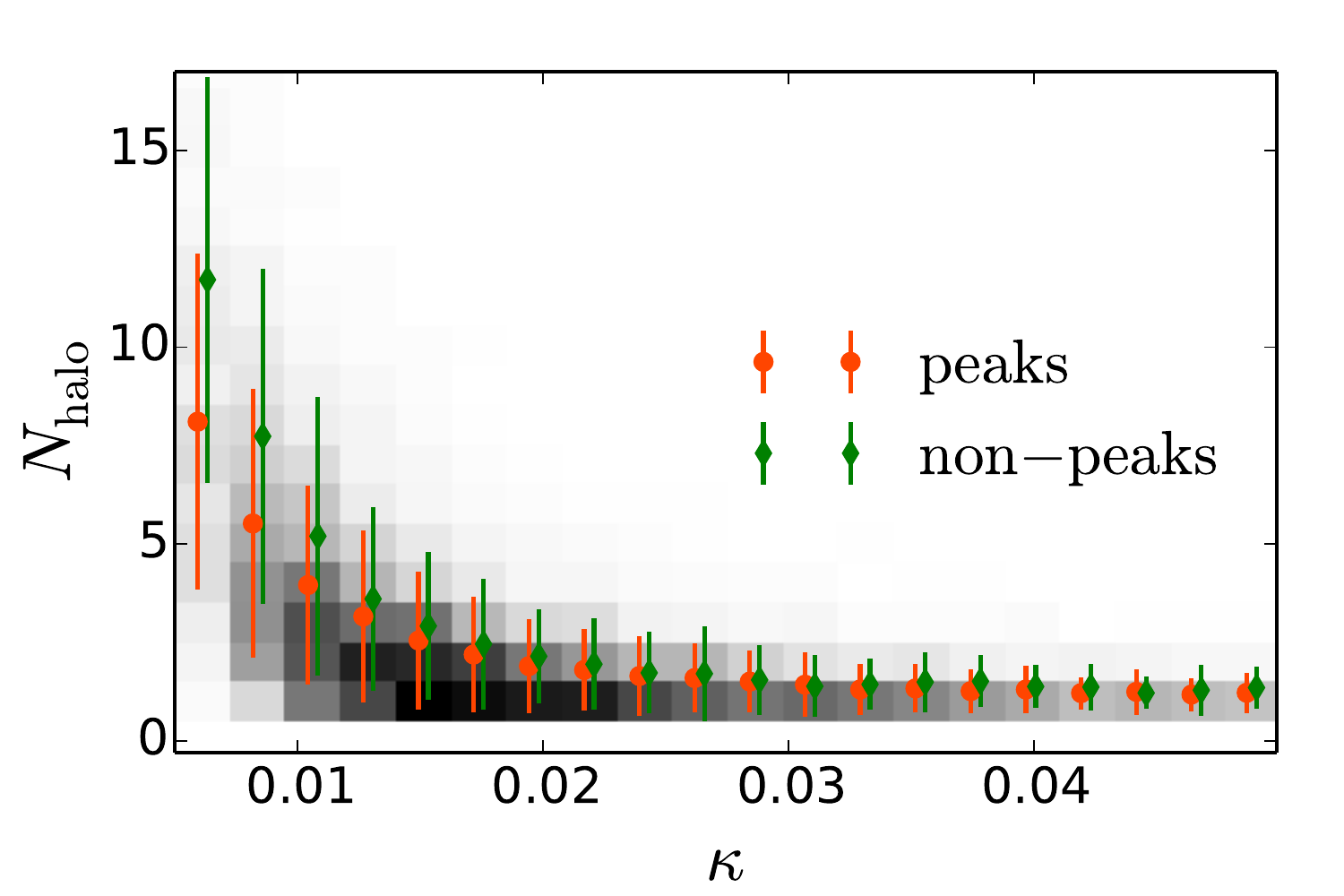}
\end{center}
\caption{\label{fig: Npeak} The number of halos needed to account for $>$50\% of the peak height (red circles) for $\kp$ peaks. We also show the same quantity, but for 10,000 random directions that are not peaks (green diamonds). The shade in the background is the two-dimensional probability distribution for all peaks. The error bar is the standard deviation of all peaks within each $\kappa$ bin. } 
\end{figure}

How many halos are needed to produce a lensing peak? To answer this question, at the position of each $\kp$ peak, we record the percentage contribution from each foreground halo within a 10 arcmin circle. We rank the halos by their contribution, starting from the highest and proceeding towards the lowest, and we add their $\kappa$ contributions. We then record the least number of halos needed to account for $>$50\% of the peak height. For this test, we use $\kp$ maps smoothed at a 1.0 arcmin scale, a relatively optimal smoothing scale for peaks~\cite{Liu2015}. In total 11,932 peaks are found. We also repeat the same exercise for 10,000 randomly selected directions that are not peaks, to study the difference between peaks and non-peaks. Our results are shown in Fig.~\ref{fig: Npeak}.

We find that high and medium peaks ($\kappa\gsim1.5\sigma_\kappa$, with $\sigma_\kappa=0.024$\footnote{We note that $\sigma_\kappa=0.024$ is smaller in $\kp$ maps than in $\kl$ maps ($\sigma_\kappa=0.036$), because the latter gain a significant contribution from shape noise, and possibly from non-halo large-scale structures.}) are mainly linked to one single halo. In contrast, low peaks are associated with several halos ($N_{\rm halo}\approx 2$ for $\kappa=0.02$ peaks, and $N_{\rm halo}\approx 8$ for $\kappa=0.0$ peaks). High $\kappa$ non-peak regions are similarly caused by one single halo, though there are very few such regions, since most pixels with $\kappa>3.5\sigma_\kappa$ are local maxima. However, to produce the low $\kappa$ non-peak regions, more halos ($\approx$ 20\%) are needed than for peaks of the same height. When we do a student's t-test, a statistical test of the null hypothesis that two samples are drawn from the same distribution, we found that while we cannot distinguish the peak and non-peak samples for $\kappa>\sigma_\kappa$, the null hypothesis is rejected at very high confidence for lower peaks (p-value $\ll$0.01). 

\begin{figure}
\begin{center}
\includegraphics[width=0.49\textwidth]{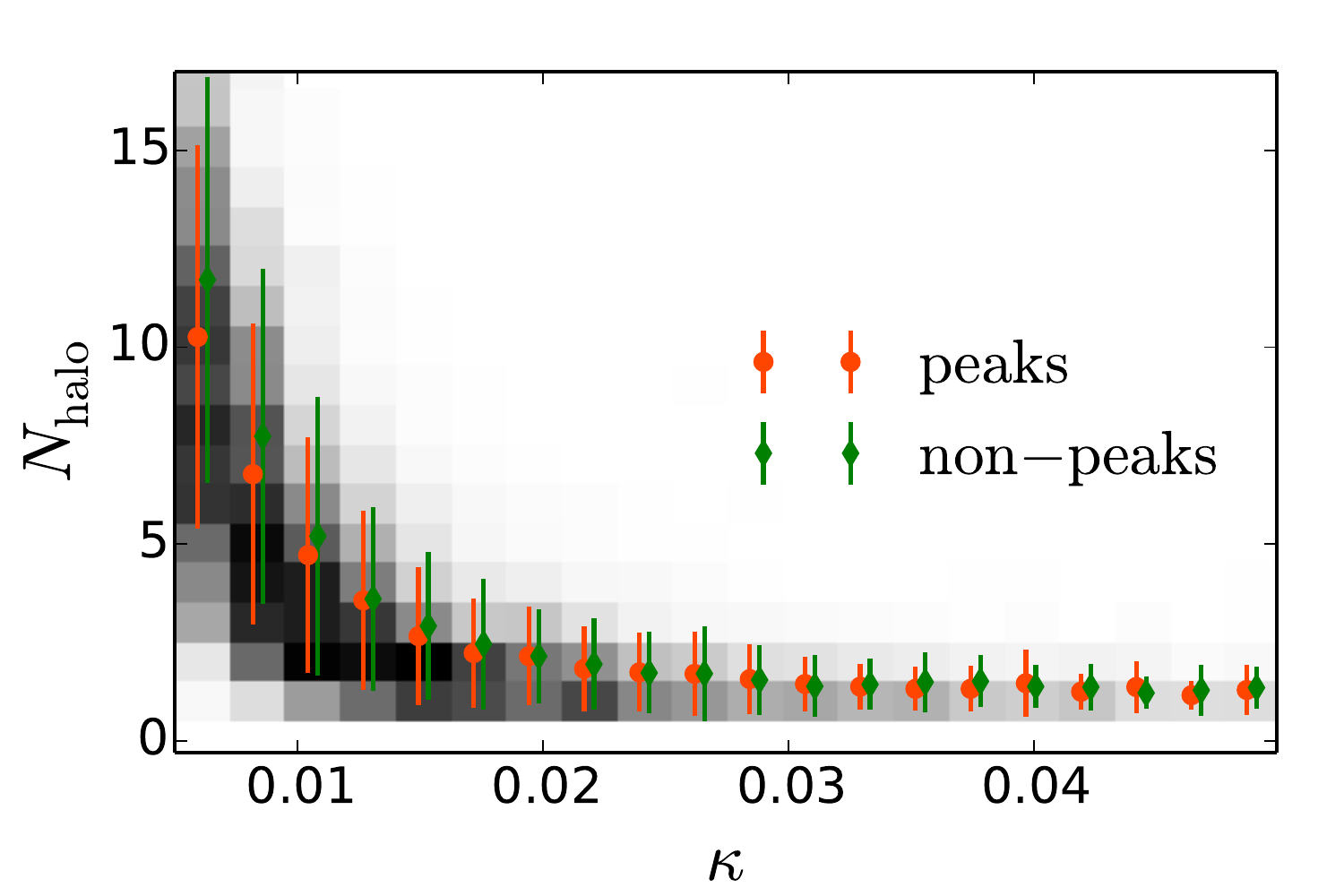}
\end{center}
\caption{\label{fig: Npeak_lens} Same as Fig.~\ref{fig: Npeak} but with the peaks extracted from $\kl$ maps. } 
\end{figure}

We repeat the same calculation for $\kl$ peaks. The results are shown in Fig.~\ref{fig: Npeak_lens}. The difference between peaks and non-peaks is smaller, due to that $\kl$ peaks are much noisier than $\kp$ peaks, receiving $\sim$ half of its variance from the galaxy shape noise. Even so, these noisy peaks still require fewer halos (2--10) than non-peaks (2--12). Our findings are consistent with results from Y11.

\begin{figure*}
\begin{center}
\includegraphics[width=\textwidth]{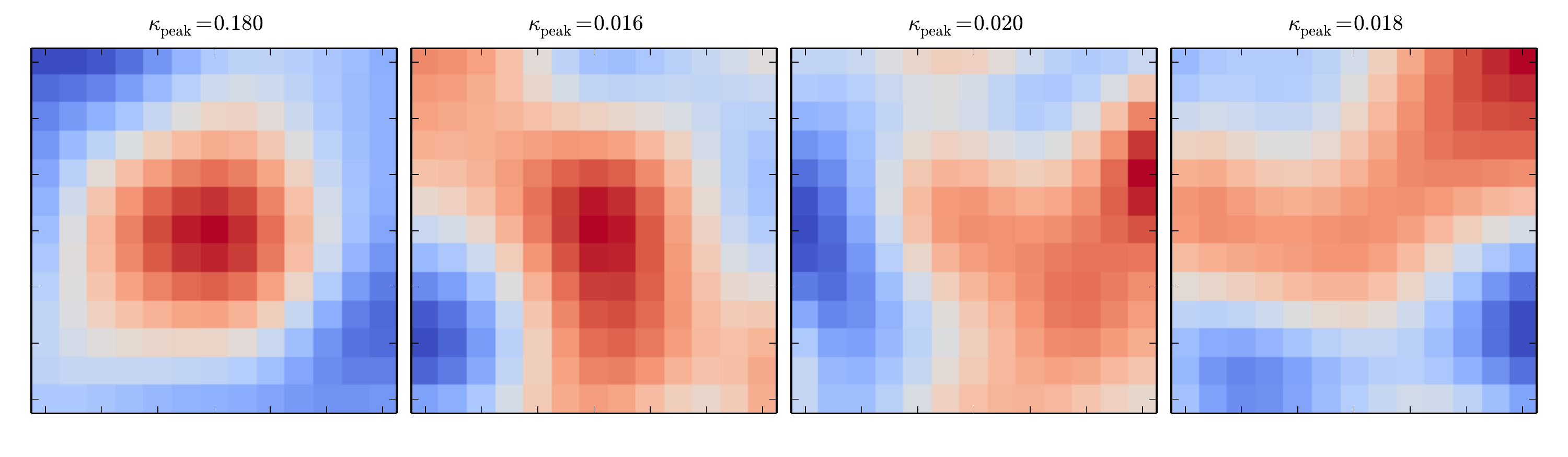}
\end{center}
\caption{\label{fig: sample_peaks} Examples of peaks in $\kl$, located at the center of each 5$\times$5 arcmin$^2$ panel. Red (blue) represents higher (lower) convergence value. The left panel shows a typical high peak with a relatively circular shape, suggesting a single massive halo. In comparison, most low peaks (showing in the other three panels) reside in complex regions where several peaks are cluttered together, or filament-like structures are seen. The peak height ($\kappa_{\rm peak}$) is marked at the top of each panel.} 
\end{figure*}

To further demonstrate the morphological difference between high and low peaks, we show examples of $\kl$ peaks with various heights in Fig.~\ref{fig: sample_peaks}. The left panel shows a typical high peak, where a clear, circular shape is seen without  other prominent structures nearby. The remaining three panels show a random selection of low peaks, where complex structures are seen, potentially due to the constellation of contributing halos. Their shapes are also less circular than the high peak, consistent with simulations~\cite{Liu2014}.

\subsection{Halo properties}

\begin{figure*}
\begin{center}
\includegraphics[width=0.9\textwidth]{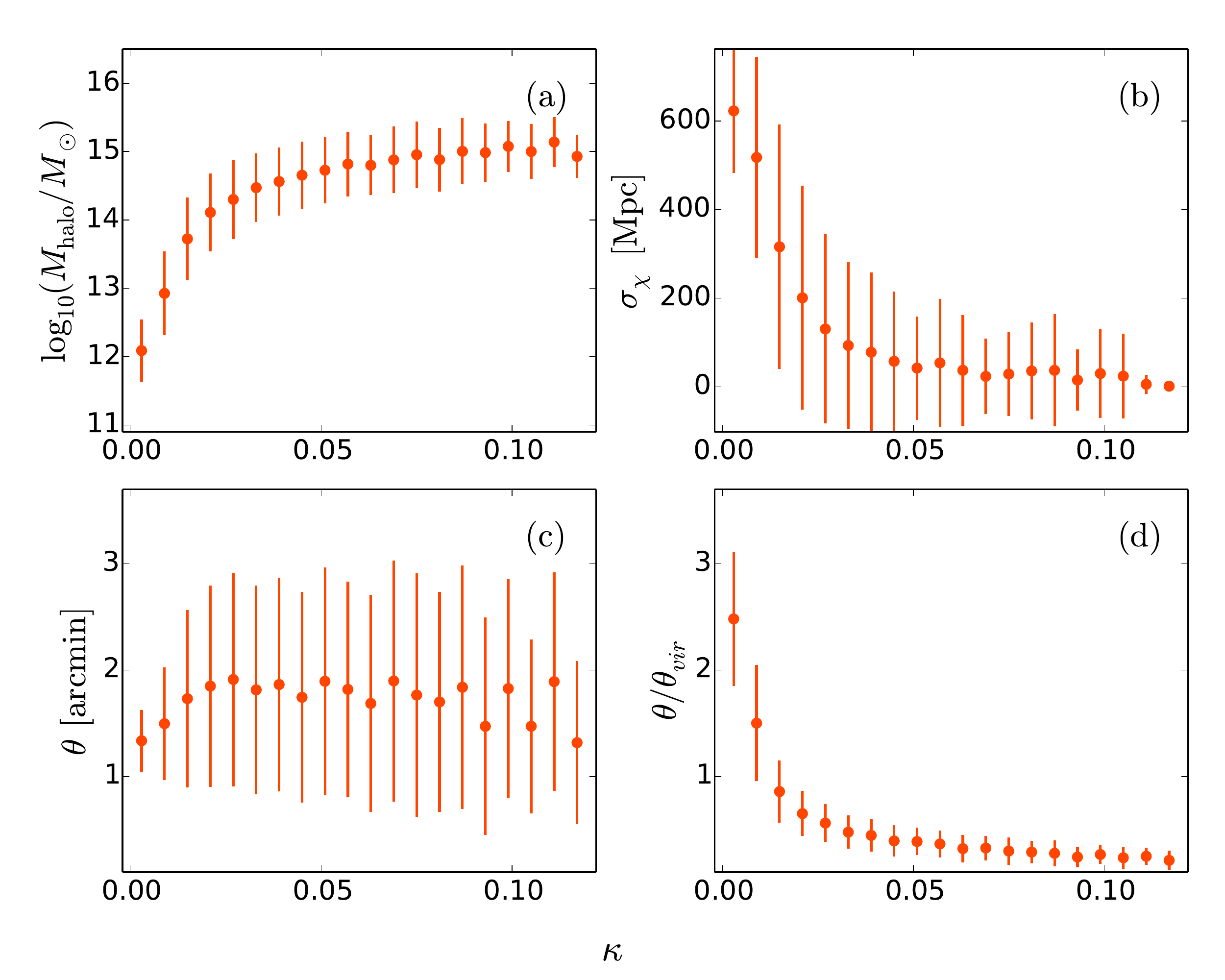}
\end{center}
\caption{\label{fig: halo_properties} Properties of halos that contribute $>$50\% of  the total convergence of $\kp$ peaks, as a function of peak height: (a) average halo mass of contributing halos, (b) comoving distance dispersion along the line of sight, (c) average angular separation between the halos and the center of the peak, in unit of arcmin, and (d) that in unit of the halo virial radius.} 
\end{figure*}

We next study properties of the halos we identified above that contribute $>$50\% of the height of $\kl$ peaks. We examine their masses, distribution along the line-of-sight, and distances from the peak position. The results are shown in Fig.~\ref{fig: halo_properties}. 

In the figure, panel (a) shows the average mass of contributing halos to each peak. It is not surprising to see that high peaks are due to one single massive halo with mass $\approx 10^{15}M_\odot$, and low peaks are associated with multiple smaller halos $\lsim10^{13}M_\odot$. 

We show in panel (b) the comoving distance dispersion of contributing halos along the line-of-sight. This is particularly interesting for low peaks that have multiple lenses---i.e. whether these lenses are clustered in one single redshift (therefore they are actually different sub-halos within a parent halo), or they are totally uncorrelated in redshift. For low peaks, the typical distance between two halos is between 200 to 600 Mpc, much larger than the typical cluster size (1--10 Mpc). If we scatter multiple halos randomly between $z=0.8$ (our mean source redshift) and $z=0.2$ (the lowest bound for lenses), we obtain a comoving distance dispersion of $\approx 550$ Mpc. Therefore, we conclude that the contributing halos for low peaks are not clustered in redshift. An interesting follow up question to this finding would be---how much of the cosmological sensitivity in low peaks is from the halo mass function (as is the case with high peaks) versus from the geometry\footnote{Since these halos are randomly distributed, their density can be sensitive to geometry through volume elements.}---we leave this to future study.

In panel (c), we show the average angular separation between these halos and the peak center. Halos are typically 1--2 arcmin away from the peak center. However, if we divide the separation ($\theta$) by the virial radius of each halo ($\theta_{\rm vir}$), we see that halos around low peaks are at 1--2.5$\theta_{\rm vir}$ distance from the peak, while high peak halos are much closer $\approx 0.25\theta_{\rm vir}$. Refs.~\cite{Yang2013, Osato2015} studied the impact of baryonic physics on peaks, which can change the matter distribution inside dark matter halos through radiative cooling and feedback. Our findings are consistent with Y11 and suggest that low peaks may be less affected by baryonic processes, which may be confined to the inner
regions of the halo (\cite{Schaller2015}, although see~\cite{Semboloni2011} who argued
that AGN feedback can impact the gas distribution out to
the virial radius and beyond).

\subsection{Beyond halos}

\begin{figure*}
\begin{center}
\includegraphics[width=\textwidth]{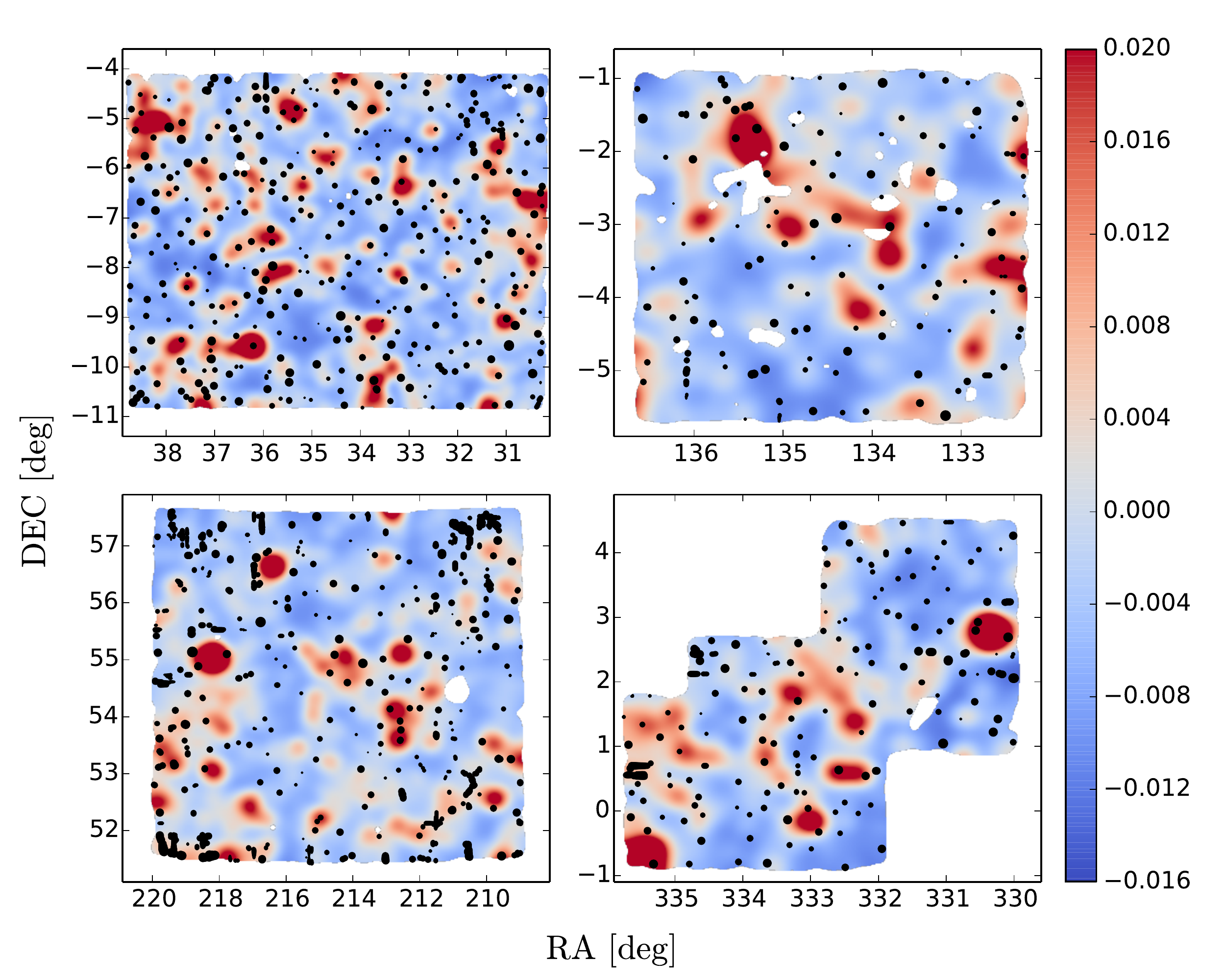}
\end{center}
\caption{\label{fig:matching_peaks_lens} Peaks (black filled circles) found in $\kl$ maps are overlaid on $\kp$ maps, for CFHTLenS W1, W2, W3, W4 fields (upper-left, upper-right, lower-left, and lower-right, respectively). Circle sizes represent the height of peaks. $\kl$ maps are smoothed with a 5.3 arcmin Gaussian window, while $\kp$ maps with a 8.9 arcmin window. These two different smoothing scales are used for the purpose of visual display. } 
\end{figure*}

Throughout this work, we have made the assumption that lensing peaks are results of the projection of one or multiple halos along the line-of-sight. Though we found strong correlation between lensing peaks and halos, it remains possible that other cosmic structures, such as filaments and walls, are responsible for lensing peaks in a non-trivial way, especially for low peaks. One such hint comes from Fig.~\ref{fig: sample_peaks}, where low peaks are often found within a complex region.  

We show the spatial distribution of lensing peaks in Fig.~\ref{fig:matching_peaks_lens}, where we overlay the $\kl$ peaks on the $\kp$ maps. We see some peaks, instead of appearing at the center of the high $\kappa$ regions implied by the halo locations, tend to cluster around these regions, such as the one at RA=216 deg, DEC=56.5 deg (lower-left panel). At RA=136 deg, DEC=$-5$ deg (upper-right panel), we see a string of peaks with no corresponding halos in the $\kp$ map. We speculate that these peaks are related to a thin filament that was not captured in $\kp$ maps~\cite{Maturi2013,Clampitt2016}. It is very unlikely that random noise produces such a coherent structure of almost a degree in size. Similarly, a cluster of peaks with no corresponding halos are also seen at RA=210 deg, DEC=$57$ deg (lower-left panel).

At this early stage of weak lensing observations, it is difficult to quantify these diffuse structures in the locations of lensing peaks with data. However, it is possible to study them in simulations. This will become important for modeling low peaks.

\section{Conclusions}\label{conclusion}

This work provides the first observational evidence for the connection between low significance lensing peaks and foreground halos, using data from the 154 deg$^2$ CFHTLenS. We produced convergence maps $\kl$ using shear measurements and $\kp$ from halo projections, and found that $\kl$ peaks are strongly correlated with these halos. 

Our main conclusions are:

1. Peak counts from CFHTLenS, despite its relatively small sky coverage compared to future surveys (100 times larger), is significantly different from that of Gaussian random fields. In particular, we see an excess of medium and high peaks ($>2\sigma_\kappa$), and a deficit of low peaks ($\lsim1\sigma_\kappa$) in $\kl$ maps, compared to the Gaussian case. These departures can encode cosmological information beyond second-order statistics.

2. High peaks are mostly associated with one single massive halo ($\approx 10^{15}M_\odot$), while low peaks are statistically associated with a constellation (i.e. not clustered in redshift) of several smaller halos ($\lsim10^{13} M_\odot$). We also found that, for low $\kp$ pixels, a random non-peak direction would require 20\% more halos to reach the same height as a peak (though this is not significant for $\kl$ peaks, which receive significant contribution from the shape noise).

3. Halos around low peaks are often smaller in mass, hence with a smaller virial radius, and low peaks are found to be offset from halos by more than their projected virial radii (compared to $\approx 0.25 \theta_{\rm vir}$ separation for high peaks), hinting that low peaks may be more immune to baryonic processes, as long as these are confined to the inner regions of the halo  (\cite{Schaller2015}, but see~\cite{Semboloni2011}).

The main caveat in this work is that, due to the difficulty in measuring halo masses directly, we derive them using the stellar-to-halo mass relation, the scatter in which can induce large uncertainties to our inferred halo masses. While other methods to infer halo masses are available, such as Halo Occupation Distribution~\cite{Zheng2005} and Halo Abundance Matching~\cite{Vale2004, Trujillo-Gomez2011}, they require additional assumptions that can induce more systematics. Our method also takes advantage of the stellar masses  already well measured by the CFHTLenS team. 

Similarly, due to the difficulty in identifying other diffuse structures, our focus is mainly on halos. It is interesting to ask if halos can account for all peaks (as assumed in all semi-analytical models so far), or if other structures, such as filaments or walls may also contribute significantly. We expect this to be more profitably investigated with simulations at least in the near future. This will be a necessary step to take, if we are to model lensing peaks (semi-)analytically down to percent level percision.

\begin{acknowledgments}
We thank the CFHTLenS team for making their data publicly available. This work has benefited from discussions with David Schiminovich, Ludo Van Waerbeke, Lance Miller, Jerry Ostriker, Colin Hill, Andrea Petri, Pedro Ferreira, and Greg Bryan. We thank Jos\'e Zorrilla, Chieh-An Lin, Masato Shirasaki, and David Spergel for comments on the manuscript.
JL is supported by an NSF Astronomy and Astrophysics Postdoctoral Fellowship under award AST-1602663. This work is supported in part by NSF grant AST-1210877 and by a ROADS award at Columbia University. This work used the Extreme Science and Engineering Discovery Environment (XSEDE), which is supported by NSF grant ACI-1053575.

\end{acknowledgments}

\bibliographystyle{physrev}
\bibliography{paper}
\end{document}